# Evaluating the Performance of ANN Prediction System at Shanghai Stock Market in the Period 21-Sep-2016 to 11-Oct-2016

**Wanjawa, Barack Wamkaya**

School of Computing and Informatics,

University of Nairobi, Kenya

wanjawawb@students.uonbi.ac.ke

**ABSTRACT**

This research evaluates the performance of an Artificial Neural Network based prediction system that was employed on the Shanghai Stock Exchange for the period 21-Sep-2016 to 11-Oct-2016. It is a follow-up to a previous paper in which the prices were predicted and published before September 21. Stock market price prediction remains an important quest for investors and researchers. This research used an Artificial Intelligence system, being an Artificial Neural Network that is feedforward multi-layer perceptron with error backpropagation for prediction, unlike other methods such as technical, fundamental or time series analysis. While these alternative methods tend to guide on trends and not the exact likely prices, neural networks on the other hand have the ability to predict the real value prices, as was done on this research. Nonetheless, determination of suitable network parameters remains a challenge in neural network design, with this research settling on a configuration of 5:21:21:1 with 80% training data or 4-year of training data as a good enough model for stock prediction, as already determined in a previous research by the author. The comparative results indicate that neural network can predict typical stock market prices with mean absolute percentage errors that are as low as 1.95% over the ten prediction instances that was studied in this research.

**Key words:**

ANN, Neural Networks, Prediction, Shanghai Stock Exchange

## 1.0 INTRODUCTION

Stock markets, and trade therein, are important for the economies of many countries. Stock trade is an investment in the financial sector, hence affects how both local and international investors are comfortable in making investment in that economy. It has been observed that stock markets react to both local and global events. In the stock market, trade in equity (stocks) tend to be quite active due to the general low entry value of trade, unlike other stock instruments such as bonds, which tend to have a relatively high entry value.

Traders at the stock exchange usually desire to maximize their investment by buying at low values and selling at a higher value to make profits (capital gain). Investors can also benefit from payment for dividends for their stock holding. Quest for getting the best deal at the stock market has led investors to desire some form of prediction method to guide their buy or sell decisions. Methods such as fundamental, technical and time series analysis are already being employed (Chen et al., 2007, Deng et al., 2011, Huang et al., 2011, Neto et al., 2009, Zhang et al., 2008), though artificial intelligence (AI) methods can also be used in prediction, with Artificial Neural Networks (ANNs) being the most preferred AI technology (NeuroAI, 2016).

Determination of optimum ANN configuration however has been a challenge, since each domain area needs the best selection of parameters suitable for the task e.g. data size and partitioning for testing and training sets, number of training cycles, number and size of nodes (input, hidden, output), network type etc. These parameters, as applicable in stock market prediction were determined in a previous research by the author (Wanjawa et al., 2014) as 5:21:21:1 with 80% data for training (4-year data). This network is still considered deep, though it only has 2-hidden layers (DL4J, 2016).

**Problem Statement:** How does an ANN-based prediction system perform, based on a 3-week advance prediction that it generates prior to the trades, as tested on some chosen stocks at the Shanghai Stock Exchange (SSE)?

## 2.0 THE ANN-BASED PREDICTION SYSTEM

The details of the model used, how it works and the experimental setup is already contained in a previous research (Wanjawa, 2016). A 4-year data set i.e. January 1, 2012 to December 31, 2015 was used for training the prediction system, while 2016 data was used for testing and also for confirming actual predictions. The experiment used 5 previous prices to predict the sixth price, and continued with such a sliding window method to predict all the prices in the period of prediction. The ANN structure is reproduced in Fig. 1 below.

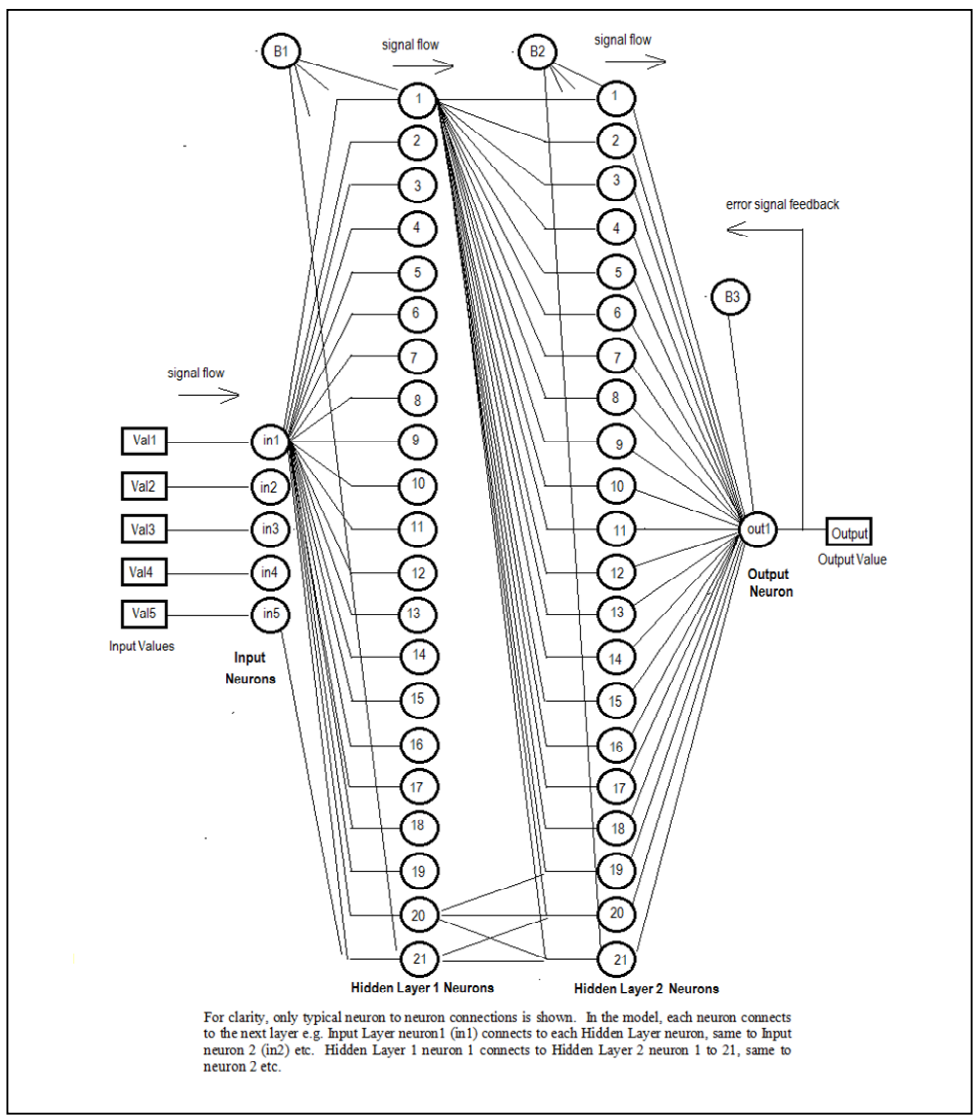

*Fig. 1* –ANN model For Stock Market Prediction (Source: Wanjawa et al., 2014)

Seven stocks were selected from the Shanghai Stock Exchange (SSE) by sampling from the numerical listing of stocks in that bourse in the stock reference range 600000 to 600100. These 7 chosen stocks are shown in Table 1 below:

*Table 1* – *List of Tested Stocks from SSE (Source: SSE, 2016)*

| Code | Short name | Short name | Full name |
|---|---|---|---|
| 600010 | 包钢股份 | BSU | Inner Mongolia BaoTou Steel Union Co.,Ltd. |
| 600015 | 华夏银行 | HUAXIA BANK | HUA XIA BANK CO., Limited |
| 600016 | 民生银行 | CMBC | CHINA MINSHENG BANK |
| 600028 | 中国石化 | Sinopec Corp. | China Petroleum and Chemical Corporation |
| 600031 | 三一重工 | SANY | SANY HEAVY INDUSTRY CO.,LTD |
| 600064 | 南京高科 | NJGK | NANJING GAOKE COMPANY LIMITED |
| 600089 | 特变电工 | TBEA | TBEA CO.,LTD. |

The previous research (Wanjawa, 2016) already published the prediction for the 15-day period from 21-Sep-2016 to 11-Oct-2016. All predicted values were generated eight trading days before September 21, 2016 i.e. on Sep. 12, 2016.

## 3.0   RESULTS

Since the previous research already published the predictions (Wanjawa, 2016), the results shown here are a comparative analysis between the predictions and the actual trades that were published by Shanghai Stock Exchange. The data used was obtained from Yahoo Finance for the 7 stocks considered i.e. 600010 (BSU, 2016), 600015 (HUAXIA BANK, 2016), 600016 (CMBC, 2016), 600028 (SINOPEC Corp., 2016), 600031 (SANY, 2016), 600064 (NJGK, 2016) and 600089 (TBEA, 2016).

It is worth noting that trade did not take place at SSE in the period Oct. 1 to Oct. 7, 2016, since this was a holiday in China (SSE, 2016b). Since the ANN model is a next day prediction system, the predictions for Oct. 3 & Oct. 4 are used for Oct. 10 and Oct. 11, since these are the 'next days' of trade after the Sep. 30 trade. The comparative results for the seven chosen stocks are shown in Table 2 to Table 8 below. The error (%) is calculated by comparing the predicted price against the 'close' price i.e. average trade price of the day.

***Table 2*** *– Comparing Actual and Predicted Prices for SSE Stock 600010*

| 600010 | Predicted | Open | High | Low | Close | Error |
|---|---|---|---|---|---|---|
| 21-Sep-16 | 2.70 | 2.78 | 2.87 | 2.78 | 2.81 | -3.91% |
| 22-Sep-16 | 2.71 | 2.82 | 2.83 | 2.79 | 2.80 | -3.21% |
| 23-Sep-16 | 2.72 | 2.81 | 2.82 | 2.79 | 2.80 | -2.86% |
| 26-Sep-16 | 2.72 | 2.80 | 2.80 | 2.74 | 2.75 | -1.09% |
| 27-Sep-16 | 2.72 | 2.75 | 2.77 | 2.74 | 2.77 | -1.81% |
| 28-Sep-16 | 2.71 | 2.77 | 2.77 | 2.74 | 2.75 | -1.45% |
| 29-Sep-16 | 2.71 | 2.75 | 2.80 | 2.75 | 2.79 | -2.87% |
| 30-Sep-16 | 2.70 | 2.78 | 2.79 | 2.77 | 2.78 | -2.88% |
| 03-Oct-16* | - | - | - | - | - | - |
| 04-Oct-16* | - | - | - | - | - | - |
| 05-Oct-16* | - | - | - | - | - | - |
| 06-Oct-16* | - | - | - | - | - | - |
| 07-Oct-16* | - | - | - | - | - | - |
| 10-Oct-16 | **2.71** | 2.78 | 2.82 | 2.78 | 2.82 | -3.90% |
| 11-Oct-16 | **2.70** | 2.82 | 2.82 | 2.80 | 2.82 | -4.26% |

*SSE was closed, due to holiday

We observe that for this stock 600010, the error swings between a low of -1.09% to a high of -4.26%. All the predictions were lower than the respective average trade price in the period under review. The error is however lower than 5% in all cases, with mean absolute percentage error (MAPE) of 2.82% for the 10 prediction instances. These results are shown graphically in Fig. 2 below.

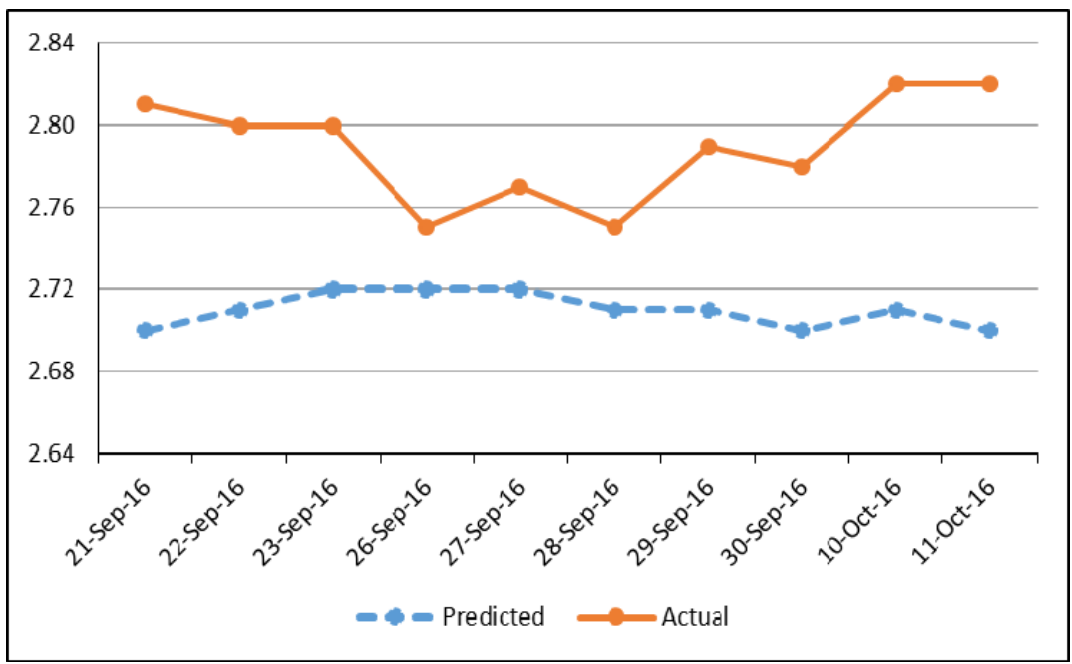

***Fig. 2*** *–Comparing Actual v/s Predicted for Stock 600010*

*Table 3* – *Comparing Actual and Predicted Prices for SSE Stock 600015*

| 600015 | Predicted | Open | High | Low | Close | Error |
|---|---|---|---|---|---|---|
| 21-Sep-16 | 10.03 | 10.05 | 10.09 | 10.03 | 10.08 | -0.50% |
| 22-Sep-16 | 10.48 | 10.10 | 10.18 | 10.08 | 10.16 | 3.15% |
| 23-Sep-16 | 10.33 | 10.16 | 10.18 | 10.14 | 10.14 | 1.87% |
| 26-Sep-16 | 10.55 | 10.13 | 10.13 | 10.02 | 10.04 | 5.08% |
| 27-Sep-16 | 10.48 | 10.02 | 10.10 | 9.98 | 10.08 | 3.97% |
| 28-Sep-16 | 10.03 | 10.09 | 10.09 | 10.01 | 10.03 | 0.00% |
| 29-Sep-16 | 10.25 | 10.03 | 10.10 | 10.03 | 10.06 | 1.89% |
| 30-Sep-16 | 10.62 | 10.04 | 10.08 | 10.02 | 10.05 | 5.67% |
| 03-Oct-16* | - | - | - | - | - | - |
| 04-Oct-16* | - | - | - | - | - | - |
| 05-Oct-16* | - | - | - | - | - | - |
| 06-Oct-16* | - | - | - | - | - | - |
| 07-Oct-16* | - | - | - | - | - | - |
| 10-Oct-16 | **10.09** | 10.05 | 10.15 | 10.05 | 10.15 | -0.59% |
| 11-Oct-16 | **10.22** | 10.17 | 10.21 | 10.12 | 10.20 | 0.20% |

*SSE was closed, due to holiday

It can be observed that for stock 600015, the error swings between a low of -0.50% to a high of +5.67%. One prediction is spot on (28-Sep-2016), two are lower than the close value, while the rest are predicted higher than the average close value. The error is lower than 6% in all cases. The MAPE over the prediction range was 2.29%. These results are also shown on Fig. 3 below.

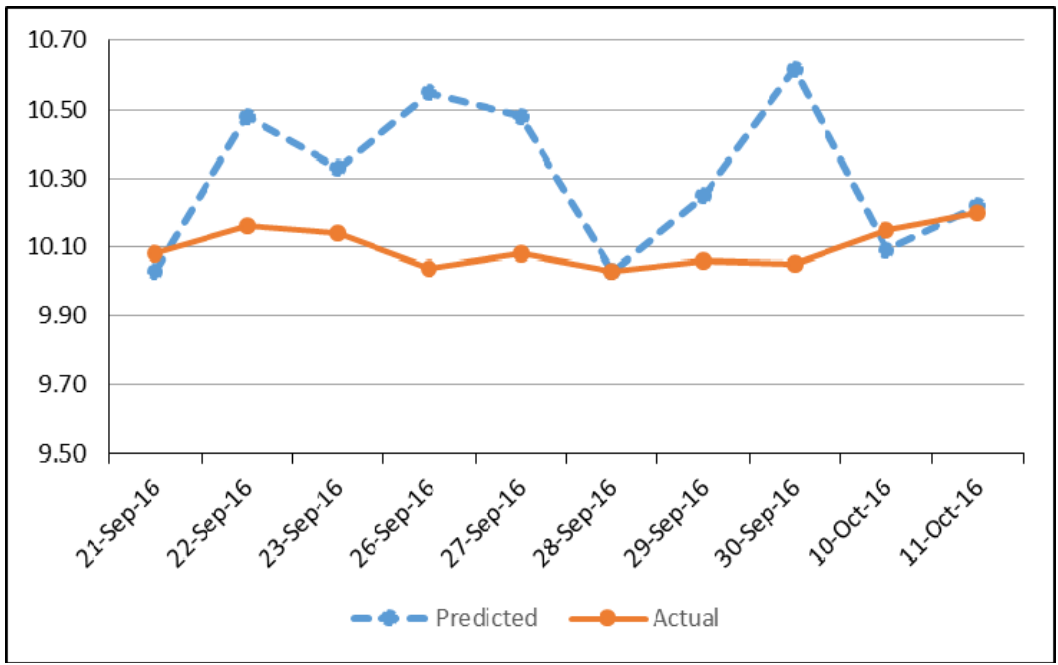

**Fig. 3** – *Comparing Actual v/s Predicted for Stock 600015*

*Table 4 – Comparing Actual and Predicted Prices for SSE Stock 600016*

| 600016 | Predicted | Open | High | Low | Close | Error |
|---|---|---|---|---|---|---|
| 21-Sep-16 | 9.07 | 9.33 | 9.34 | 9.30 | 9.33 | -2.79% |
| 22-Sep-16 | 8.89 | 9.35 | 9.40 | 9.32 | 9.37 | -5.12% |
| 23-Sep-16 | 8.70 | 9.38 | 9.39 | 9.34 | 9.38 | -7.25% |
| 26-Sep-16 | 8.63 | 9.28 | 9.35 | 9.25 | 9.31 | -7.30% |
| 27-Sep-16 | 8.69 | 9.30 | 9.32 | 9.26 | 9.31 | -6.66% |
| 28-Sep-16 | 8.80 | 9.31 | 9.31 | 9.16 | 9.20 | -4.35% |
| 29-Sep-16 | 8.79 | 9.21 | 9.25 | 9.20 | 9.23 | -4.77% |
| 30-Sep-16 | 8.69 | 9.23 | 9.29 | 9.23 | 9.26 | -6.16% |
| 03-Oct-16* | - | - | - | - | - | - |
| 04-Oct-16* | - | - | - | - | - | - |
| 05-Oct-16* | - | - | - | - | - | - |
| 06-Oct-16* | - | - | - | - | - | - |
| 07-Oct-16* | - | - | - | - | - | - |
| 10-Oct-16 | **8.59** | 9.26 | 9.36 | 9.26 | 9.33 | -7.93% |
| 11-Oct-16 | **8.49** | 9.32 | 9.34 | 9.28 | 9.31 | -8.81% |

*SSE was closed, due to holiday

The Table 4 results for stock 600016 indicate that the prediction error ranges between -2.79% and -8.81%, with a MAPE of 6.11%. All predictions tended to be lower than the actual average close for the respective date of trade. The results are shown graphically in Fig. 4 below.

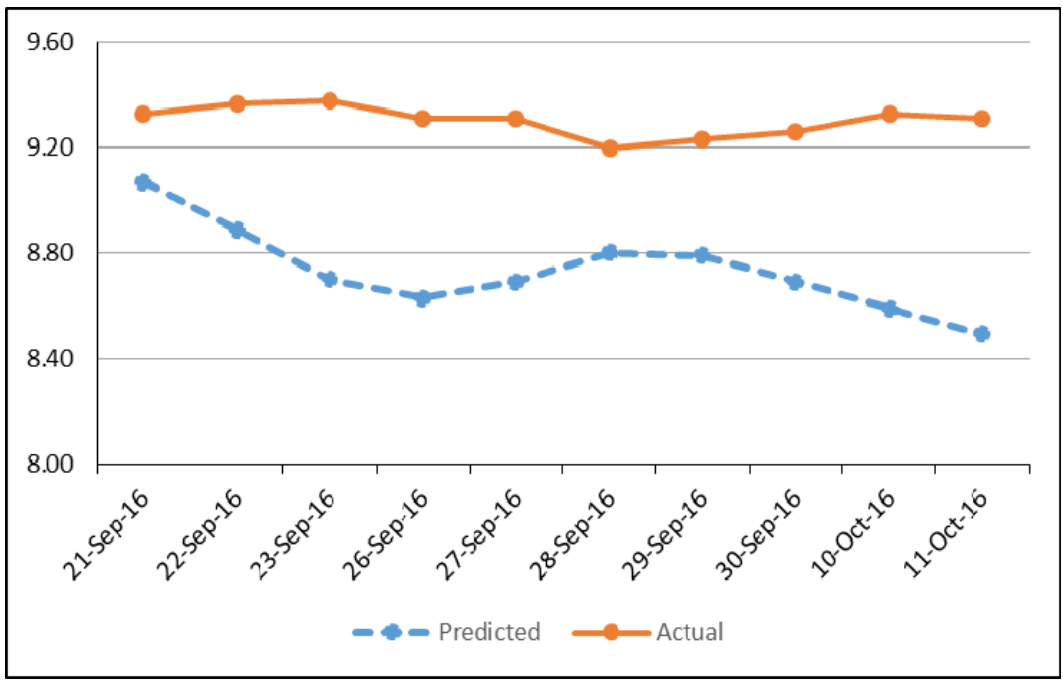

**Fig. 4** – *Comparing Actual v/s Predicted for Stock 600016*

*Table 5 – Comparing Actual and Predicted Prices for SSE Stock 600028*

| 600028 | Predicted | Open | High | Low | Close | Error |
|---|---|---|---|---|---|---|
| 21-Sep-16 | 4.93 | 4.78 | 4.79 | 4.74 | 4.78 | 3.14% |
| 22-Sep-16 | 4.93 | 4.79 | 4.86 | 4.78 | 4.84 | 1.86% |
| 23-Sep-16 | 4.93 | 4.85 | 4.87 | 4.84 | 4.85 | 1.65% |
| 26-Sep-16 | 4.93 | 4.84 | 4.85 | 4.78 | 4.80 | 2.71% |
| 27-Sep-16 | 4.93 | 4.79 | 4.82 | 4.77 | 4.81 | 2.49% |
| 28-Sep-16 | 4.93 | 4.80 | 4.80 | 4.76 | 4.77 | 3.35% |
| 29-Sep-16 | 4.93 | 4.81 | 4.86 | 4.80 | 4.85 | 1.65% |
| 30-Sep-16 | 4.93 | 4.83 | 4.86 | 4.82 | 4.86 | 1.44% |
| 03-Oct-16* | - | - | - | - | - | - |
| 04-Oct-16* | - | - | - | - | - | - |
| 05-Oct-16* | - | - | - | - | - | - |
| 06-Oct-16* | - | - | - | - | - | - |
| 07-Oct-16* | - | - | - | - | - | - |
| 10-Oct-16 | **4.93** | 4.86 | 4.97 | 4.86 | 4.95 | -0.40% |
| 11-Oct-16 | **4.93** | 4.96 | 4.99 | 4.94 | 4.97 | -0.80% |

*SSE was closed, due to holiday

The results above, for stock 600028, show that the prediction errors ranged between -0.40% and +3.35%. It is also quite interesting that the prediction system predicted the same figure over the duration of the prediction. However, the error was lower than 3.4% in all cases, with MAPE obtained for this stock being the lowest of all the seven tested stocks at 1.95% over the ten predictions.

*Table 6 – Comparing Actual and Predicted Prices for SSE Stock 600031*

| 600031 | Predicted | Open | High | Low | Close | Error |
|---|---|---|---|---|---|---|
| 21-Sep-16 | 5.08 | 5.63 | 5.68 | 5.60 | 5.60 | -9.29% |
| 22-Sep-16 | 5.17 | 5.61 | 5.65 | 5.58 | 5.60 | -7.68% |
| 23-Sep-16 | 5.23 | 5.60 | 5.65 | 5.59 | 5.60 | -6.61% |
| 26-Sep-16 | 5.40 | 5.58 | 5.65 | 5.44 | 5.45 | -0.92% |
| 27-Sep-16 | 5.38 | 5.45 | 5.48 | 5.37 | 5.45 | -1.28% |
| 28-Sep-16 | 5.28 | 5.47 | 5.48 | 5.40 | 5.41 | -2.40% |
| 29-Sep-16 | 5.10 | 5.41 | 5.47 | 5.40 | 5.43 | -6.08% |
| 30-Sep-16 | 5.15 | 5.41 | 5.48 | 5.41 | 5.47 | -5.85% |
| 03-Oct-16* | - | - | - | - | - | - |
| 04-Oct-16* | - | - | - | - | - | - |
| 05-Oct-16* | - | - | - | - | - | - |
| 06-Oct-16* | - | - | - | - | - | - |
| 07-Oct-16* | - | - | - | - | - | - |
| 10-Oct-16 | **5.21** | 5.47 | 5.57 | 5.45 | 5.52 | -5.62% |
| 11-Oct-16 | **5.35** | 5.52 | 5.65 | 5.52 | 5.59 | -4.29% |

*SSE was closed, due to holiday

For stock 600031, we observe that the prediction error swings between -0.92% and +9.29%. All prediction were values lower than the average closing price. The prediction error was less than 9.3% in the prediction period, with MAPE being 5.00%.

*Table 7* – *Comparing Actual and Predicted Prices for SSE Stock 600064*

| 600064 | Predicted | Open | High | Low | Close | Error |
|---|---|---|---|---|---|---|
| 21-Sep-16 | 16.20 | 17.17 | 17.22 | 17.00 | 17.17 | -5.65% |
| 22-Sep-16 | 16.08 | 17.24 | 17.49 | 17.17 | 17.18 | -6.40% |
| 23-Sep-16 | 15.98 | 17.20 | 17.36 | 17.15 | 17.17 | -6.93% |
| 26-Sep-16 | 15.87 | 17.10 | 17.11 | 16.67 | 16.69 | -4.91% |
| 27-Sep-16 | 15.78 | 16.68 | 16.96 | 16.61 | 16.95 | -6.90% |
| 28-Sep-16 | 15.72 | 17.02 | 17.26 | 16.95 | 17.15 | -8.34% |
| 29-Sep-16 | 15.67 | 17.15 | 17.35 | 17.10 | 17.27 | -9.26% |
| 30-Sep-16 | 15.63 | 17.25 | 17.55 | 17.25 | 17.51 | -10.74% |
| 03-Oct-16* | - | - | - | - | - | - |
| 04-Oct-16* | - | - | - | - | - | - |
| 05-Oct-16* | - | - | - | - | - | - |
| 06-Oct-16* | - | - | - | - | - | - |
| 07-Oct-16* | - | - | - | - | - | - |
| 10-Oct-16 | **15.61** | 17.51 | 17.51 | 16.81 | 17.11 | -8.77% |
| 11-Oct-16 | **15.61** | 17.07 | 17.26 | 17.00 | 17.11 | -8.77% |

*SSE was closed, due to holiday

Looking at the data on Table 7, we note that the error swings between a low of -4.91% to a high of -10.74% for this stock 600064. This was the single stock where the prediction values tended to be a bit far from the actual average closing price, with all predictions being relatively lower in value than the actual trades. One of the predicted prices had an error of over 10%. This prediction range also realized the highest error (MAPE) of 7.67%.

*Table 8* – *Comparing Actual and Predicted Prices for SSE Stock 600089*

| 600089 | Predicted | Open | High | Low | Close | Error |
|---|---|---|---|---|---|---|
| 21-Sep-16 | 8.91 | 8.82 | 8.85 | 8.80 | 8.83 | 0.91% |
| 22-Sep-16 | 8.89 | 8.87 | 8.91 | 8.82 | 8.84 | 0.57% |
| 23-Sep-16 | 8.89 | 8.84 | 8.85 | 8.77 | 8.79 | 1.14% |
| 26-Sep-16 | 8.89 | 8.78 | 8.79 | 8.66 | 8.67 | 2.54% |
| 27-Sep-16 | 8.88 | 8.63 | 8.66 | 8.56 | 8.65 | 2.66% |
| 28-Sep-16 | 8.88 | 8.67 | 8.67 | 8.55 | 8.58 | 3.50% |
| 29-Sep-16 | 8.87 | 8.63 | 8.63 | 8.58 | 8.59 | 3.26% |
| 30-Sep-16 | 8.87 | 8.60 | 8.64 | 8.58 | 8.62 | 2.90% |
| 03-Oct-16* | - | - | - | - | - | - |
| 04-Oct-16* | - | - | - | - | - | - |
| 05-Oct-16* | - | - | - | - | - | - |
| 06-Oct-16* | - | - | - | - | - | - |
| 07-Oct-16* | - | - | - | - | - | - |
| 10-Oct-16 | **8.87** | 8.62 | 8.78 | 8.62 | 8.77 | 1.14% |
| 11-Oct-16 | **8.87** | 8.78 | 8.79 | 8.73 | 8.78 | 1.03% |

*SSE was closed, due to holiday

We observed that for stock 600089, the error swings between +0.57% and +3.50%, with all predicted prices tending to be relatively higher than the actual average traded figures. The ANN prediction system also tended to predict same values on days that follow each other. The error was lower than 3.5% in all cases, with a MAPE of 1.96% over the ten predicted values.

We can make general observations that the ANN prediction system, which had predicted all these values on Sep. 12, 2016, generated a generally good prediction method, with almost all predictions made for the seven different stocks falling below an error rate of 10% from the actual closing price. It is worth noting that the prediction system had to first predict the eight preceding stock prices, before predicting the ones for the research period of Sep. 21 to Oct. 11, 2016. The prediction range was therefore quite wide – from Sep. 12 all the way to Oct. 11, 2016.

The trading rules of SSE (SSE, 2015 & SSE, 2016c), spell out that a swing of 10% price movement is the maximum allowed at any day of trade, based on the previous day's closing price. This would mean that the prediction system predicted prices that were practically tradable at the SSE. However, the traders determine the actual price swings on any given days, and this swing can be very low e.g. sometimes even 0% (constant price). It is therefore essential that the prediction system not only conforms to the maximum swing rule, but should also be as close to the actual average market price as possible i.e. the actual supply and demand prices. This supply/demand prices may not necessarily obey the 10% swing e.g. for Stock 600010, the swing observed was between 0% and 1.8% in the period under consideration. Additionally, the prediction system should be able to track the trend of the price movement itself (up and down over time). The ANN prediction system was not only able to predict with low error (absolute on any trade day and also MAPE over the prediction period), but also provide an acceptable price trend movement.

## 4.0    CONCLUSION

This research tends to make us believe that Artificial Neural Network (ANN) prediction systems achieve high prediction accuracy in typical application domains, such as stock market price prediction due to their exploitation of parallel computing to learn from training data. After gaining the knowledge from learning, the ANN system can use this intelligence on new data that it has not yet been exposed to in achieving such a regression task. A carefully designed ANN system, such as one of configuration 5:21:21:1, with 4-year training data can be practical for prediction. This configuration was carefully chosen by a previous research and is proving applicable to prediction of typical stock exchanges trades such as that of Shanghai Stock Exchange (SSE).

In this research, the ANN system was trained on the 4-year data for the period 2012 to 2015, then used for prediction of stock prices for trades done in 2016. The research also predicted the prices of seven(7) stocks of the SSE for the period Sep. 12 to Sep. 20 in advance (as at Sep. 12, 2016), then went on to continue the prediction for values from Sep. 21 all the way to Oct. 11, 2016. All these predictions were published by Sep. 12, 2016 and it was a matter of wait and see if the predictions would turn out true on the respective dates of trade.

Considering the period Sep. 21 to Oct. 11, 2016, the results show that the ANN-based prediction system was able to predict within the 10% price swing limit for all the 7-stocks in consideration (70 values) as dictated by one of the trading rules at the bourse i.e. allowing only a maximum of 10% difference between the previous close with the current day's trading values. The individual daily errors were also quite narrow e.g. between +0.91% and +3.50% for Stock 600089 for the 10 predicted instances.

This research confirms that a fairly simple deep ANN configuration of only 2-hidden layers in depth can achieve good prediction results, based on the mean absolute percentage error (MAPE) realized over the prediction period for the seven test stocks that ranged from 1.95% for Stock 600028 to 7.67% for Stock 600064 considering the 10 predictions for each stock. This fairly simple design makes us believe that deep networks must not necessarily be complex, but should be carefully chosen and experimented depending on the task at hand.

The future of deep ANN remains promising, especially for prediction in the stock market domain. More work need to be done in studying the effects of deepening the network beyond the 2-layers, tweaking the number of nodes and even incorporating other stock exchange parameters such as traded volumes and market sentiment on to the prediction system with a view of seeing if the predictions obtains have significance improvements over the current system in terms of absolute daily predictions and even the MAPE over a range of predictions.